\documentclass[draft]{agujournal2019}
\usepackage{url} 
\usepackage{lineno}
\usepackage[inline]{trackchanges} 
\usepackage{soul}
\usepackage{natbib}
\usepackage{multirow}
\draftfalse

\journalname{Geophysical Research Letters}

\begin{document}

%
%


\title{Detectability of Forced ENSO Changes under Global Warming: Insights from the Recharge Oscillator}

%
%




\authors{Sooman Han\affil{1}, Jérôme Vialard\affil{2}, Alexey V. Fedorov\affil{1,2}, and Soong-Ki Kim\affil{1}}


\affiliation{1}{Department of Earth and Planetary Sciences, Yale University, New Haven, CT, USA}
\affiliation{2}{LOCEAN-IPSL, Sorbonne Université-CNRS-IRD-MNHN, Paris, France}




\correspondingauthor{Sooman Han and Jérôme Vialard}
{sooman.han@yale.edu; jerome.vialard@locean.ipsl.fr}



\begin{keypoints}
\item Recharge Oscillator (RO) perfect-model experiments test whether slow, forced changes can be detected from finite ENSO time series.
\item Observed ENSO amplitude, period, and asymmetry trends since 1955 are unlikely but cannot be ruled out under internal variability alone.
\item Our study quantifies the detectability of RO-diagnosed forced ENSO changes, helping their interpretation in climate-model ensembles.
\end{keypoints}

%
%

%
%

\begin{abstract} 
We use perfect-model, large-ensemble nonlinear Recharge Oscillator (RO) simulations to quantify ENSO internal variability and the detectability of forced changes in ENSO characteristics. Fixed-parameter simulations show that linear trends in ENSO standard deviation, period, and skewness as large as those observed since 1955 can arise from internal variability in 10\% of simulations. RO parameters estimated from single realizations exhibit spurious drifts without forcing, underscoring the need for ensembles. Using 100-member ensembles, comparable to the largest climate-model ensembles, we identify detectable parameter trends. For ENSO amplitude, the detectability thresholds for forced changes in underlying processes are 15\% per century for stochastic forcing, 25\% per century for basin adjustment, and 50\% per century for the Bjerknes feedback. For ENSO period, the detectability threshold for forced changes in the underlying recharge--discharge processes and delayed oceanic feedback is 15\% per century. These results provide a testbed for interpreting RO-diagnosed ENSO changes in climate-model ensembles.
\end{abstract}

\section*{Plain Language Summary} 

The El Niño–Southern Oscillation (ENSO) is a major source of year-to-year climate variability worldwide. Its characteristics, including amplitude, period, and asymmetry, vary from decade to decade. These changes may result from natural internal variability or from gradual changes in the tropical Pacific climate driven by rising greenhouse gas concentrations. Here, we use a simple, physically based ENSO model—the Recharge Oscillator (RO)—to determine whether externally forced changes in ENSO can be distinguished from natural variability. In the RO framework, ENSO dynamics are represented by a small set of parameters describing key physical processes. Simulations with fixed parameters show that internal variability alone can generate multidecadal changes in ENSO properties comparable to those observed since 1955. The observed trends fall near the 90th percentile of internally generated variability, indicating that trends of comparable magnitude can arise naturally with 10\% of probability. Climate models increasingly provide large ensembles of up to 100 members, allowing natural variability to be sampled more effectively. Using 100-member ensembles, we quantify how large changes in RO parameters must be for their effects to become distinguishable from internal variability. Our results provide guidance for identifying and interpreting forced ENSO changes in climate-model ensembles.

%
%

%


%
%
%
%

\section{Introduction}

The El Niño–Southern Oscillation (ENSO) is the leading mode of interannual global climate variability \citep{Trenberth20review}. Although ENSO arises from air–sea interactions in the tropical Pacific, its remote impacts shape rainfall, drought, and climate extremes worldwide, with major societal consequences \citep{McPhaden20reviewch1}. ENSO properties, such as amplitude and period, depend on the mean ocean–atmosphere state, which is being altered by anthropogenic climate change \citep{Fedorov20review}. For example, greenhouse warming can strengthen upper-ocean thermal stratification \citep{Capotondi12} and reduce background trade winds \citep{Heede21}. In fact, most climate models project a long-term weakening of the Pacific trade winds, equatorial zonal SST gradient, and upwelling in response to increasing greenhouse gases, even though observed trends over the past 40 years have been opposite \citep{Wills22, Heede23b, Watanabe24}. Therefore, how these mean-state changes affect ENSO remains a central question, given the potentially massive economic losses associated with future ENSO strengthening \citep{Callahan23}.

Observed ENSO amplitude increased from the mid 20th to the early 21st century \citep{Fedorov20review}, a trend reproduced by some, but not all, climate models \citep{Cai23, Maher23, Coquereau25}. Many projections indicate a further increase by the late 21st century \citep{Cai21, Heede23b}; however, large ensembles suggest that the forced response may be nonmonotonic \citep{Maher23}, with ENSO amplitude potentially weakening under sufficiently strong warming \citep{Callahan21, Peng24, Tuckman25}. Whereas \citet{IPCC21} assessed projected changes in ENSO amplitude as highly uncertain because of model biases and uncertain mean-state responses, more recent work has challenged this assessment by identifying a robust increase across four IPCC emissions scenarios \citep{Cai22}. Overall, no consensus has emerged regarding how ENSO will change in a warming climate. Compounding this uncertainty, internal variability can produce pronounced interdecadal modulation of ENSO even in the absence of external forcing \citep[e.g.,][]{Wittenberg09,Zheng18,Ng21}. Robust estimates of ENSO properties therefore require large ensembles \citep{Planton24}, which are essential for separating forced changes from internal variability \citep[e.g.,][]{Vega-Westhoff17,Maher18,Milinski20,Maher23,Coquereau25}.

Uncertainties in projected ENSO changes amid internal variability highlight the need for complementary, process-oriented diagnostics that can relate changes in ENSO characteristics to changes in the underlying dynamics. One such approach is using the Recharge Oscillator (RO) framework, a highly reduced but physically-based representation of ENSO dynamics in which a relatively small set of parameters summarizes integrated physical processes and feedbacks \citep{Jin97a, Jin97b, Jin20review, Vialard25}. These include the Bjerknes feedback \citep{Bjerknes69}—the air–sea interaction loop that allows equatorial Pacific SST anomalies to self-amplify—stochastic forcing associated with Westerly Wind Bursts (WWB) \citep{Fedorov02, Lengaigne04, Eisenman05, Yu20, Yu22}, the delayed negative feedback from ocean adjustment \citep{Wyrtki75, Jin97a, Fedorov09}, and nonlinearities that help shape ENSO asymmetry and extremes \citep{An20review, Stuivenvolt-Allen25b}. In this framework, a given parameter set encodes both ENSO dynamics and its resulting statistical properties \citep{Levine16, Lu18, Kim20, Jin20review, Kim21}. In practice, RO parameters are estimated by fitting the RO model to time series obtained from observations or GCM simulations, raising a key question: can mean-state-driven parameter changes be robustly inferred from finite records in the presence of internal variability?

Motivated by these issues, in this study we address two main questions:(1) how large are internally generated ENSO variations in the RO framework, and how likely are the observed 1955–2025 ENSO trends under internal variability alone? (2) how large must forced RO parameter trends and their associated ENSO changes be to become detectable in finite ensembles? To answer these questions, we use a perfect-model framework in which synthetic ENSO time series are generated with either fixed RO parameters or prescribed parameter trends. We focus on 100-member ensembles, representative of the largest current single-model ensembles \citep{Maher19, Coquereau25}, and compare their apparent trends with 5,000-member ensemble means, which provide an estimate of the underlying forced response. Section~2 describes the RO formulation and experiment setups. Section~3 quantifies internally generated ENSO trends in the fixed-parameter case and compares them with observed trends since 1955. Section~4 uses prescribed parameter-change experiments to identify which RO parameter trends, and associated ENSO-property changes, can be detected with 100-member ensembles. Section~5 discusses the implications for interpreting RO-diagnosed ENSO changes in observations and climate-model ensembles.

\section{Methods}

The RO configuration in this study follows \citet{Han26}, who formulated a stable, WWB-driven nonlinear system and systematically explored its parameter space to identify the configuration that best reproduces the observed characteristics of ENSO, including amplitude, skewness, autocorrelation, power spectrum, frequency, and seasonality (Figure~S1):

\begin{linenomath*}
\begin{equation}
\frac{dT}{dt} = RT + F_1 h + b_TT^2 - c_TT^3 +\sigma_T w_T(1+BH(T)T) 
\end{equation}

\begin{equation}
R=R_0-R_a\cos(\omega_at-\phi)
\end{equation}

\begin{equation}
\frac{dh}{dt} = -\varepsilon h - F_2 T -b_hT^2 + \sigma_h w_h
\end{equation}
\end{linenomath*}

\noindent Here, $T$ and $h$ denote dimensionless Sea Surface Temperature Anomaly (SSTA) over the Niño~3 region (150°W–90°W, 5°S–5°N) and Ocean Heat Content Anomaly (OHCA) over the western Pacific (120°E–180°E, 5°S–5°N), each normalized by its own standard deviation. This normalization gives both the linear ($R$, $F_1$, $\varepsilon$, $F_2$) and nonlinear ($b_T$, $c_T$, $b_h$) parameters units of month$^{-1}$. Equations~1 and 3 describe the coupled evolution of SSTA and OHCA under unit-variance white Gaussian noise forcing ($w_T$ and $w_h$), scaled by $\sigma_T$ and $\sigma_h$ \citep{Jin97a, Burgers05}. Equation~2 represents the seasonal modulation of SSTA growth, with $R_0$ and $R_a$ denoting the mean and seasonal components, $\omega_a=2\pi/12$ the annual frequency, and $\phi$ so that $R$ reaches its maximum in boreal fall \citep{Kim21}. The parameter $B>0$ controls multiplicative noise, and $H(T)$ is the Heaviside step-function, equal to 1 for $T>0$ and 0 otherwise \citep{Levine10,Levine16}. Details are given in Text~S1.

We consider six distinct RO simulation families: one fixed-parameter reference simulation, and five forced-trend simulations, in which $R_0$, $\varepsilon$, $\sigma_T$, $F_1$, and $F_2$ are varied linearly in time (Table~S1). We impose trends of 5–50\% per century in $R_0$, and 5–30\% per century in the other parameters. This range is designed to span the forced ENSO-amplitude trends diagnosed in climate-model large ensembles, which remain within approximately 5–50\% per century in \citet{Maher23}. The signs of the imposed trends are motivated by previous studies: increases in $R_0$ by enhanced thermocline feedback \citep{Zhu24} and stronger wind response to SST anomalies \citep{Stuivenvolt-Allen25a}; decreases in $\varepsilon$ and $F_2$ by the meridional widening of ENSO-related wind-stress anomalies \citep{Stuivenvolt-Allen25a, Stuecker25}; increases in $F_1$ by stronger thermocline influence on SST anomalies under enhanced thermal stratification \citep{Iwakiri25}; and increases in $\sigma_T$ by stronger stochastic forcing in a warmer climate \citep{ferrett19, Heede23b, Liang24}. These incremental forcing levels allow us to determine how large a forced change must be before it becomes distinguishable from internal variability. Detectability is evaluated for three ENSO metrics—standard deviation, period, and skewness—as well as for RO parameters estimated from synthetic monthly SSTA and OHCA time series using the RO tendency equations \citep{Burgers05}. The formal definition of detectability is provided in Section~4.

For each case, we generate 5,000 ensemble members with different initial conditions and stochastic noise realizations, following \citet{Jin07}. Each simulation lasts 350 years; the first 50 years are discarded as spin-up and the remaining 300 years are analyzed. Parameters remain fixed throughout the fixed-parameter simulations. In the forced-change experiments, they are fixed during spin-up and then varied linearly over the following 300 years. The stochastic equations are integrated with an Euler–Maruyama scheme using a 0.01-month time step.

Text~S2 explains why regression-based parameter estimates are expected to display a systematic bias, as first noted by \citet{Weeks25}. The bias arises from the finite-difference approximation applied to monthly data, i.e. $dT/dt \approx (T_{n+1}-T_n)/\Delta t$, with $\Delta t=1$ month and $T_n$ the SSTA at month $n$. Figure~S2 summarizes parameter estimates from 5,000 members for fixed-parameter simulations under perfect-model assumptions, i.e. using the correct model structure and prescribed parameters in the regression. Numbers indicate the systematic bias of the ensemble mean relative to the true parameter values. Most parameters are recovered within $\sim$10\%, except for $R_0$ and $c_T$, for which even small absolute biases translate into large relative errors. Because our focus is on internal variability and temporal changes, rather than absolute parameter values, this baseline bias is removed in the subsequent analyses.

\section{Internal Variability Complicates the Attribution of Observed Interdecadal ENSO Changes}

Figure~1(a,b) compare the observed ENSO time series from the HadISST Niño~3 index (150$^\circ$--90$^\circ$W, 5$^\circ$S--5$^\circ$N) \citep{Rayner03} with a synthetic fixed-parameter RO simulation, illustrating their close resemblance. Figure~1(c,e,g) show the observed interdecadal modulation of ENSO standard deviation, period estimated from the first minimum of the lagged autocorrelation \citep{Jiang21}, and skewness, computed from 40-year moving windows applied to the time series in Figure~1(a). Over the 70-year period since 1955, the linear trend in standard deviation increased by $\sim$28\% from 0.72 to 0.92, the diagnosed period lengthened by $\sim$25\% from 34.9 to 43.5 months, and skewness increased by $\sim$108\% from 0.45 to 0.93. We also fit the RO equations to the same sliding 40-year observational windows using ORAS5 western Pacific heat content (120$^\circ$E--180$^\circ$, 5$^\circ$S--5$^\circ$N) \citep{Zuo19}, and overlay the resulting BJ index, $\mathrm{BJ}=(R-\varepsilon)/2$; Wyrtki index, $\mathrm{Wyrtki}=\sqrt{F_1F_2-(R+\varepsilon)^2/4}$; and $b_T$ in Figure~1(c,e,g). These quantities serve as proxies for ENSO amplitude \citep{Jin06,Kim20}, period \citep{Lu18}, and skewness \citep{Kim20}, respectively. These fitted quantities generally vary with the observed ENSO metrics, which might be interpreted as evidence that mean-state changes have altered ENSO dynamics and thereby driven the observed ENSO changes.

Figure~1(d,f,h) shows the same analysis applied to the RO simulation in Figure~1(b). Although this simulation uses time-invariant parameters, stochastic forcing generates substantial internally-driven interdecadal fluctuations in ENSO properties. Figure~S3 shows that the remaining fitted parameters likewise exhibit substantial apparent drifts. In principle, all fitted parameters should remain constant and match the dashed “true” values. In practice, however, the fitted BJ index, Wyrtki index, and $b_T$ exhibit clear interdecadal modulation and co-vary with internally driven fluctuations in ENSO amplitude, period, and skewness. This indicates that an RO fit to a single realization tends to track decadal ENSO modulations, regardless of whether they arise from external forcing or internal stochastic variability. Consequently, the agreement between observed ENSO changes and inferred RO parameter changes largely reflects internally driven changes in ENSO characteristics and cannot, by itself, be interpreted as evidence of forced changes in ENSO dynamics.

Although internally driven variations in ENSO properties can be large, they may only rarely produce sustained upward trends over a 70-year period such as those observed since 1955. We therefore ask how unlikely the observed trends in ENSO standard deviation, period, and skewness are under internal variability alone. To this end, Figure~1(i,j,k) compare the observed trends with distributions of 70-year trends from fixed-parameter RO simulations. The observed trends lie near the 90th percentile of the internally generated distributions. Thus, positive trends of comparable or greater magnitude can arise from internal variability alone about 10\% of the time. The 95th-percentile thresholds are much larger—43\%, 41\%, and 194\% over 70 years for standard deviation, period, and skewness, respectively—confirming that the observed trends remain below the 95\% confidence level.

Thus, although the observed trends in ENSO properties would be quite rare according to the RO simulations, their origin in internal variability cannot be excluded. This highlights the difficulty of detecting externally forced ENSO changes from a single realization and motivates the use of ensemble-mean approaches that suppress internal variability.

\section{Detectability of Forced RO Parameters and ENSO Metric Changes}

We next assess whether forced changes in RO parameters and the associated ENSO responses can be detected in 100-member ensemble means. We choose 100 members because this strongly reduces internal variability while remaining representative of the largest climate-model ensembles, providing an idealized test of what could be diagnosed from such ensembles \citep[e.g.,][]{Maher18, Maher23}

Figure~2 shows experiments in which one RO parameter varies linearly by 20\% per century while all others remain fixed; the 5,000-member ensemble mean defines the true forced response. Note that the retrieved forced trends for RO parameters do not exactly match the imposed changes of 20\% per century because of the finite-difference errors described in Text~S2, but the discrepancies are generally less than 3\% and therefore acceptable. The signs of the induced amplitude changes are consistent with analytical expectations for the standard deviation in the linear RO framework \citep{Kim20}: increasing $R_0$ or $\sigma_T$, or decreasing $\varepsilon$ or $F_2$, increases the ENSO standard deviation, whereas increasing $F_1$ decreases it. The sensitivity ranks from largest to smallest as $\sigma_T$, $\varepsilon$, $F_2$, $R_0$, and $F_1$. ENSO period is controlled primarily by $F_1$ and $F_2$, whereas the effects of $R_0$ and $\varepsilon$ are secondary, consistent with the Wyrtki-index formulation \citep{Lu18}. Notably, changes in skewness are also induced, most strongly by $\varepsilon$, followed by $F_1$, and more weakly by $\sigma_T$ and $F_2$, even though the nonlinear parameters remain fixed. This likely reflects interactions between the linear and nonlinear terms, although the underlying mechanism is not quantified here.

Figure~2 also reveals that substantial internal variability remains even after averaging 100 members. As in the observational analysis of Section~3, detectability depends on whether this residual variability can produce sustained trends as large as the forced response. We therefore apply the same percentile-based approach to 100-member ensemble means: each percentile ranks the 5,000-member “true” trend against trends generated by internal variability alone in the fixed-parameter simulations. For example, a 20\% per century increase in $R_0$ yields a fitted $R_0$ trend at the 82.5th percentile, implying that internal variability alone could generate such a trend 17.5\% of the time. Its induced ENSO-amplitude response has a similar 16.1\% probability of being internally generated. Thus, neither the fitted $R_0$ trend nor its amplitude impact is significant at the 95\% confidence level in a 100-member ensemble. By contrast, 20\% per century trends in $\varepsilon$, $\sigma_T$, $F_1$, and $F_2$, as well as their main ENSO impacts—amplitude changes for $\varepsilon$, $\sigma_T$, and $F_2$, and period changes for $F_1$ and $F_2$—are statistically significant.

Because detectability varies across RO parameters and ENSO properties, Figure~3 investigates the imposed trend magnitude required for detection. It shows the signal-to-noise ratio $\eta=f_{5000}/\sigma_{100}$, where $f_{5000}$ is the forced trend determined as the 5,000-member ensemble mean and $\sigma_{100}$ is the standard deviation of trends from 10,000 Monte Carlo samples of 100-member ensemble means. Because $\eta$ increases approximately linearly with the imposed trend, the threshold $|\eta|>2$ provides an estimate of the forcing required for detection at approximately the 95\% confidence level. The first column asks how large a forced RO-parameter trend must be to be reliably recovered by fitting a 100-member ensemble, whereas the remaining columns ask how large that trend must be to induce a detectable change in ENSO standard deviation, period, or skewness.

Trends in $\varepsilon$, $\sigma_T$, $F_1$, and $F_2$ become detectable at imposed magnitudes of 5–10\% per century, whereas $R_0$ requires an increase of 45\%, substantially higher possibly because its detection is complicated by multiplicative noise. Detecting the resulting changes in ENSO properties generally requires larger forced trends than detecting the parameter changes themselves. The respective detectability thresholds are 45\% and 50\% per century for $R_0$ and ENSO amplitude, 10\% and 25\% for $\varepsilon$ and ENSO amplitude, 5\% and 15\% for $\sigma_T$ and ENSO amplitude, and 10\% and 15\% and 5\% and 15\% for $F_1$ and $F_2$, respectively, and the associated ENSO-period changes. Detectable changes in ENSO skewness are induced by 20\% per century in $\varepsilon$ and 25\% per century in $F_1$. Therefore, these idealized experiments indicate that 100-member ensembles can detect certain dynamical changes and their associated ENSO trends at thresholds of approximately 15\% per century for some parameters and higher values for others, broadly comparable to the magnitudes reported in existing climate-model large ensembles \citep{Maher23}.

\section{Discussions and conclusion}

Using Monte Carlo ensembles of 100 members drawn from 5,000 RO simulations based on \citet{Han26}, we quantify internal variability in key ENSO properties and determine how large externally forced changes must be to emerge from it. Consistent with GCM studies showing that recent observed ENSO amplitude changes may lie near the upper tail of internal variability \citep{Vega-Westhoff17, Cai23}, we find that the observed 1955--2025 trends in ENSO standard deviation, period, and skewness lie near the 90th percentile of internally generated trends. Firm attribution to anthropogenic forcing is therefore difficult from observations alone, and the observed changes likely reflect both internal variability and forced response. This is consistent with the strongly stochastic nature of ENSO \citep{Vijayeta18}, which complicates the attribution of observed ENSO changes \citep{Bodai25}. This is also consistent with large-ensemble historical GCM simulations, some of which reproduce the observed strengthening of ENSO since the mid-20th century, whereas many do not, depending on several factors, including ensemble size, ENSO sensitivity to radiative forcing, and the magnitude of internal variability in a given model \citep{Maher23, Coquereau25}.

When the RO model is fitted to observations, its parameters exhibit pronounced interdecadal variability that could appear to explain observed changes in ENSO properties. However, our perfect-model experiments show that similar erroneous parameter variations arise when the RO model is fitted to pseudo-observations generated with fixed parameters. This indicates that single-realization parameter estimates are strongly contaminated by internal variability and may misinterpret noise-driven fluctuations as forced dynamical changes. Because the fitting procedure is designed to reproduce the properties of a particular time series, apparent temporal changes in fitted RO parameters do not, by themselves, demonstrate that ENSO dynamics have genuinely changed. Several previous studies have interpreted fitted RO parameters changes as indicators of mean-state induced dynamical changes \citep{Geng20, Kim20, Crespo22, Crespo23}. Although those studies also presented independent evidence for consistent mean-state changes, such interpretations should still be treated with caution.

An ensemble framework is essential for suppressing internal variability and attributing diagnosed parameter changes to genuine dynamical evolution. Although substantial internal variability remains in 100-member ensemble means, it rarely manifests as sustained trends over approximately 100-year periods, allowing sufficiently large forced changes to emerge. Detectability nevertheless varies considerably across ENSO metrics and RO parameters. The impacts of residual internal variability are smallest for ENSO period, larger for standard deviation, and largest for skewness. For RO parameters, these impacts increase from $\sigma_T$ to $F_2$, $\varepsilon$, $F_1$, and $R_0$. Consistent with this hierarchy, trends of 5–15\% per century in $\sigma_T$, $\varepsilon$, $F_1$, and $F_2$ are detectable in 100-member ensembles, whereas $R_0$ requires a much larger trend of 45\% per century. This has important implications for interpreting forced ENSO changes. Noise-driven changes in ENSO amplitude are relatively easy to detect, whereas changes in the Bjerknes feedback are much harder to identify, which may help explain why the Bjerknes feedback is often a poor indicator of ENSO amplitude in GCM analyses \citep{ferrett19, Heede23b}. Although we did not assess nonlinear parameter trends, our results suggest that changes in skewness—often used to infer ENSO asymmetry and extreme El Niño behavior \citep{Cai20review, Bayr24, Liu24}—are especially difficult to attribute, because skewness has large internal variability and is also sensitive to changes in linear parameters such as $\varepsilon$ and $F_2$.

Our conclusions also appear robust to the choice of reference RO configuration. Figure~S6 compares internal variability from four observationally constrained fixed-parameter configurations, including the \citet{Han26} baseline, a \citet{Kim21}-based configuration, and two perturbed variants within reported uncertainty ranges. The resulting internal variability ranges are broadly similar, indicating that our main conclusions do not depend on the specific \citet{Han26} parameter set. Understanding which processes and RO parameters control the remaining differences in ENSO interdecadal variability remains an important topic for future work.

Forced ENSO-amplitude trends diagnosed in GCM large ensembles are typically of order 5–50\% per century \citep{Maher23}. Our idealized experiments suggest that 100-member ensembles can detect most linear RO parameter trends of this magnitude, with thresholds near 5–15\% per century, except for Bjerknes-feedback changes, which require much larger trends of 45\% per century. The RO framework can therefore help diagnose forced changes in ENSO dynamics from the largest available climate-model ensembles. Such ensembles remain rare, however: only three of the 14 single model large ensembles analyzed by \citet{Maher23} exceed 50 members, and only two approach 100 members. Even fewer large ensembles consider different future climate scenarios \citep{Coquereau25}. This highlights both the value of existing large ensembles for understanding ENSO’s response to climate change and the need for more. The ensemble size required to detect changes in ENSO dynamics depends on the parameter considered, the trend magnitude, and the length of the analysis window; this should be explored systematically in future work.

Our results emphasize the importance of detecting slow, sustained changes as internally generated variability in ENSO characteristics remains substantial even in 100-member means, but only rarely produces coherent trends over $\sim$100-year windows. Here, we diagnose such changes in a conservative way: RO parameters are fitted separately for each ensemble member and each time window, and continuous trends are then estimated from these window-by-window fits. This procedure is useful for quantifying detectability, but it does not exploit the full information contained in the ensemble. An ensemble Kalman filter or related data-assimilation approaches could instead fit a slowly varying RO parameter set directly to the full ensemble evolution, jointly using the ensemble mean and spread. Such methods may reduce overfitting to internal variability and improve the detection of forced ENSO-dynamical changes in smaller ensembles.

\section*{Conflict of Interest Statement}
The authors have no conflicts of interest to disclose.

\section*{Open Research Section}
HadISST data \citep{Rayner03} are available at \url{https://www.psl.noaa.gov/data/timeseries/month/DS/Nino3}. ORAS5 \citep{Zuo19} data are available via \citet{ORAS5data}. The RO simulation code is archived on Zenodo \citep{Han25ROCode}. The simulation data and figure-generation code are archived separately on Zenodo \citep{HanROdata26}.



\acknowledgments 
The authors acknowledge the use of ChatGPT-5.4 for editorial assistance in improving manuscript readability. A. V. F. was supported by NASA (80NSSC21K0558) and DOE (DE-SC0023134). J. V. was supported by the ARISE project (ANR-18-CE01-0012) and the ``Tropico'' project funded by the LEFE program of the Institut des Sciences de l'Univers (INSU). S.-K. K. was supported by the Climate \& Global Change Postdoctoral Fellowship.

%
%

\bibliography{ENSO}

\begin{figure}[htbp]
  \centering
  \includegraphics[width=1.0\textwidth]{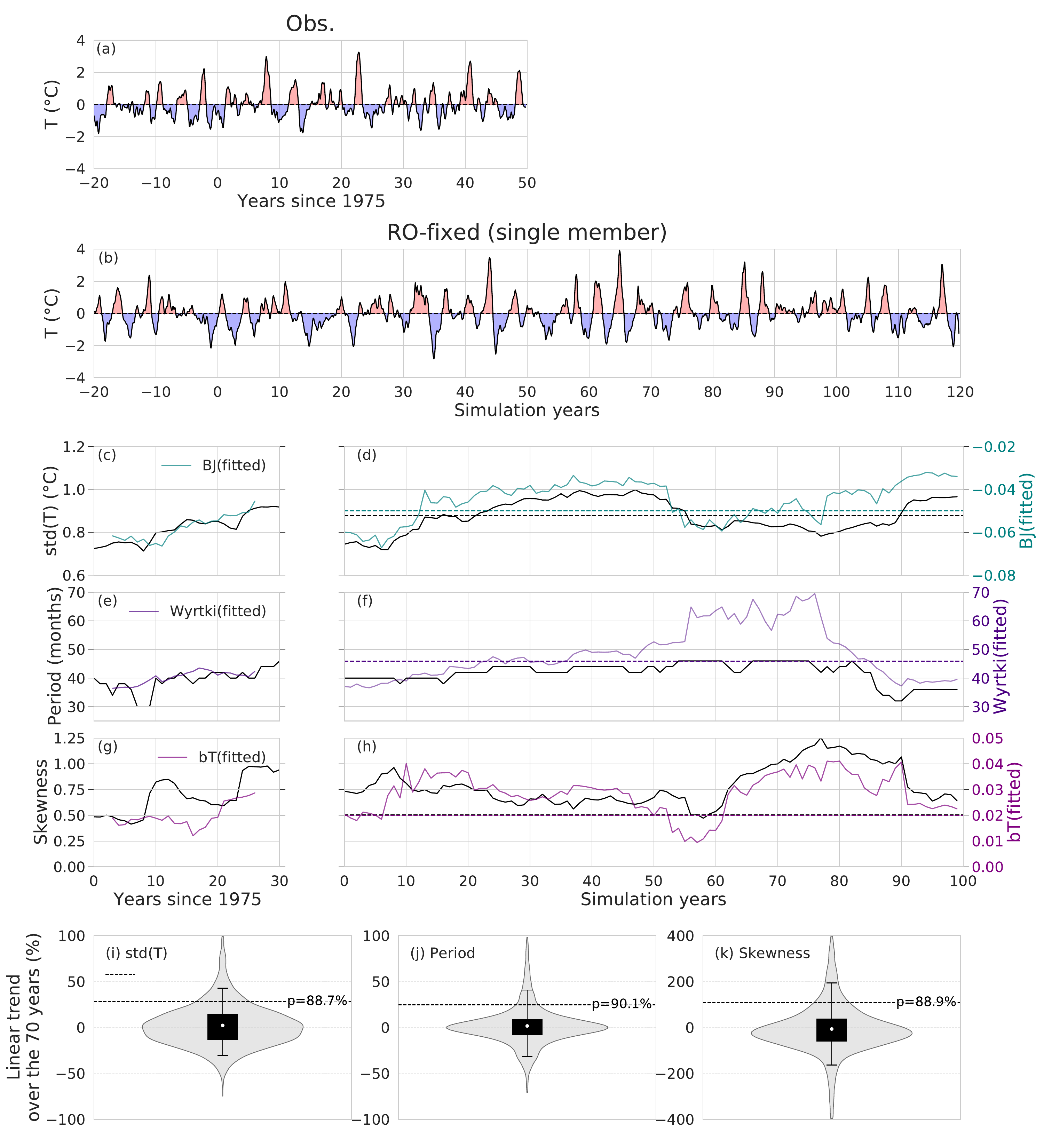}
  \caption{ENSO interdecadal modulation in observations and fixed-parameter RO simulations. (a,b) The Niño~3 index from observations and from a single experiment within a fixed-parameter RO ensemble based on the configuration of \citet{Han26}, respectively. (c,e,g) Niño~3 index standard deviation, period, and skewness computed using a 40-year moving windows applied to (a), with Year~0 denoting the 1955–1995 window mean. Regression-based estimates of the BJ index, Wyrtki index, and $b_T$ are also shown. (d,f,h) As in (c,e,g), but for the synthetic time series in (b); dashed lines indicate the 5,000-member mean metrics and the true parameter values. ENSO modulation comparable to that observed, and variations in fitted parameter values, emerge from internal variability alone, even when true parameters are fixed. (i–k) Violin plots of 70-year linear trends generated by internal variability alone in fixed-parameter RO ensemble, compared with the observed trends. White dots, boxes, and whiskers denote the mean, 25th–75th percentiles, and 5th–95th percentiles, respectively. The observed ENSO trends lie near the 90th percentile of internally generated trends, indicating that they are unlikely, yet still possible, under internal variability alone.
  }
\end{figure}

\begin{figure}[htbp]
  \centering
  \includegraphics[width=1.0\textwidth]{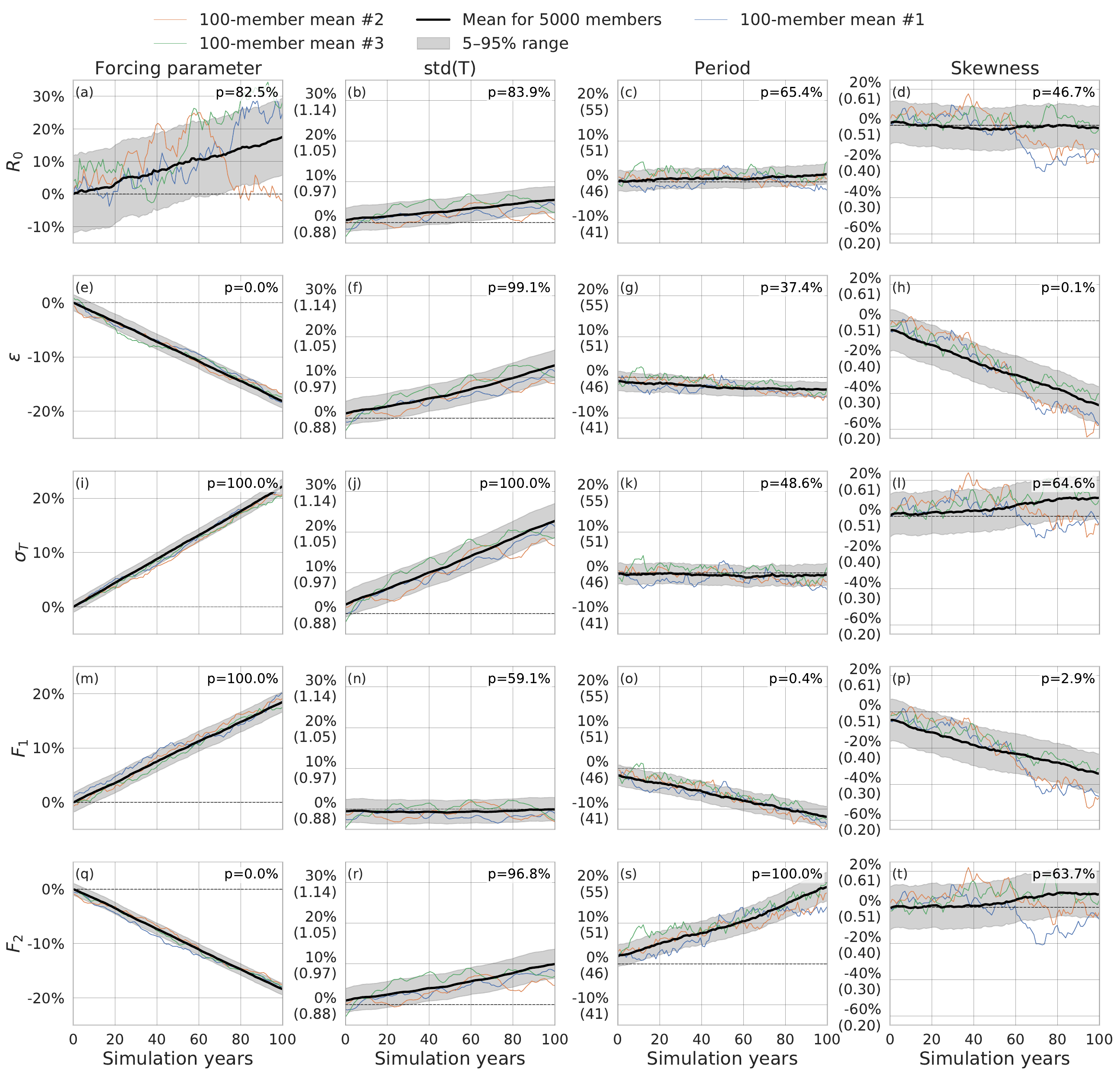}
  \caption{
  Forced changes and internal variability in fitted RO parameters and ENSO metrics for 100-member ensemble means. Rows correspond to RO simulations with imposed parameter trends of 20\% per century: increases in $R_0$, $\sigma_T$, and $F_1$ in rows 1, 3, and 4, and decreases in $\varepsilon$ and $F_2$ in rows 2 and 5. The first column shows the fitted trend in each perturbed parameter, while the remaining columns show the resulting trends in ENSO standard deviation, period, and skewness. Thick black lines denote the 5,000-member ensemble mean, treated as the forced response. Shading shows the 95\% range of 100-member ensemble means estimated from 10,000 Monte Carlo samples, and colored lines show three representative samples. The reported percentiles indicate where the true forced response falls within the distribution of 100-member-mean trends generated by internal variability alone in fixed-parameter simulations (Figures~S4 and S5). Trends in $\varepsilon$, $\sigma_T$, and $F_2$, together with their effects on ENSO amplitude, are statistically significant at the 95\% confidence level, whereas the trend in $R_0$ and its effect on amplitude are not. Changes in $F_1$ and $F_2$, along with the associated responses in ENSO period, are also statistically significant. In addition, trends in $\varepsilon$ and $F_1$ produce large changes in ENSO skewness.
  }
\end{figure}


\begin{figure}[htbp]
  \centering
  \includegraphics[width=1.0\textwidth]{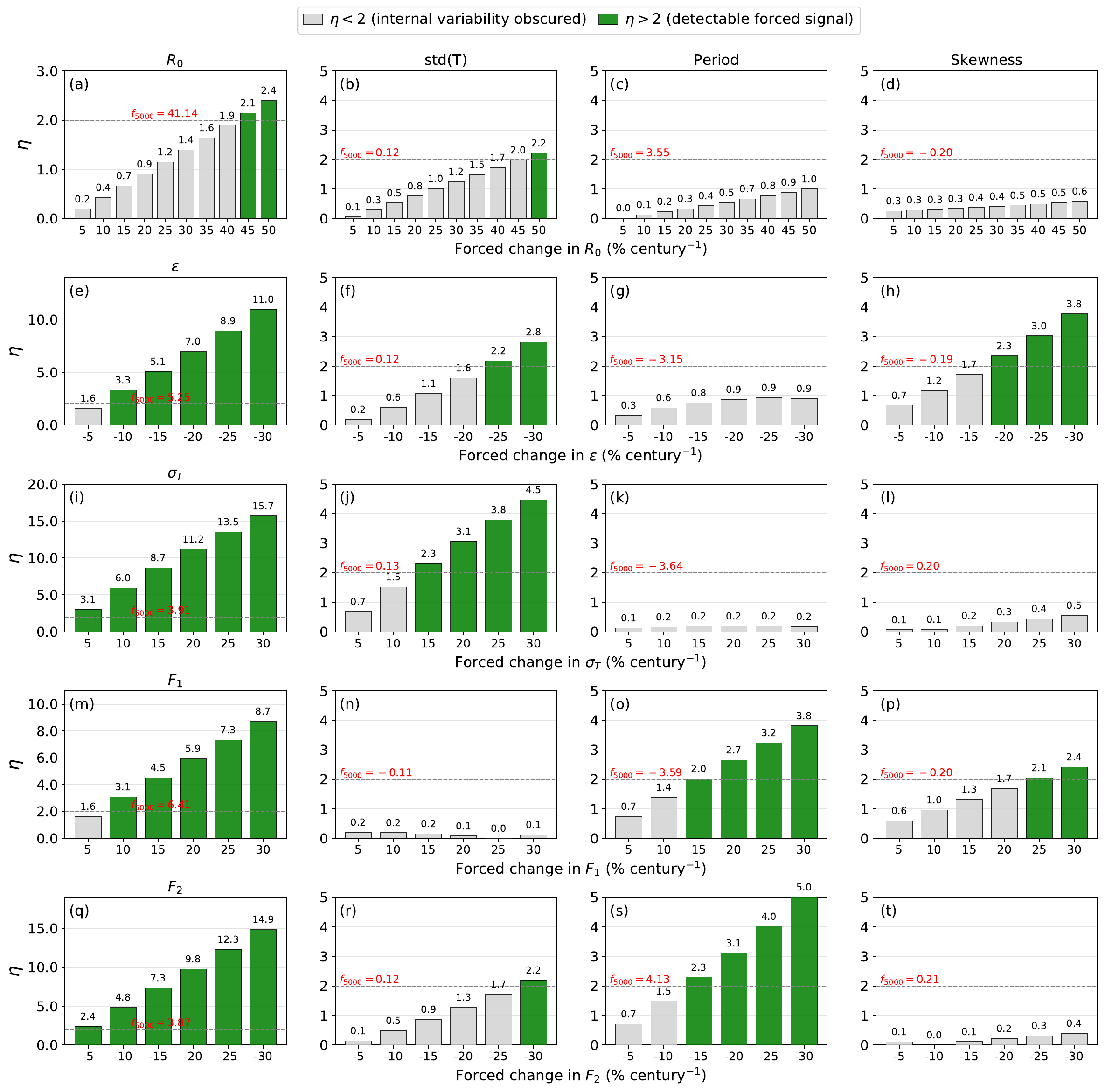}
  \caption{
  Detectability of forced changes in RO parameters and ENSO metrics using 100-member ensemble means. Rows show simulations with progressively larger imposed trends in $R_0$, $\varepsilon$, $\sigma_T$, $F_1$, and $F_2$. The first column shows how large a true trend in an RO parameter must be to be recovered by fitting a 100-member ensemble, whereas the remaining columns show how large that trend must be to induce a detectable change in ENSO standard deviation, period, or skewness. Detectability is measured by the signal-to-noise ratio $\eta=f_{5000}/\sigma_{100}$, where $f_{5000}$ is the forced trend estimated from the 5,000-member ensemble mean and $\sigma_{100}$ is the standard deviation of trends from 10,000 Monte Carlo samples of 100-member ensemble means. Values $|\eta|>2$ fall outside the approximate 95\% internal variability range and are considered detectable; red labels indicate $f_{5000}$ at the detectability threshold, expressed as percentage changes for RO parameters and absolute changes for ENSO metrics. Fitted parameter trends become detectable at imposed changes of approximately 45\% per century for $R_0$, 10\% per century for $\varepsilon$ and $F_1$, and 5\% per century for $\sigma_T$ and $F_2$, indicating that changes in the Bjerknes feedback are the most difficult to detect. Changes in ENSO properties generally require larger imposed trends than their controlling parameters—for example, 15\% versus 5\% per century for $\sigma_T$ and ENSO amplitude, and 15\% versus 10\% for $F_1$ and ENSO period.
  }
\end{figure}

%
%
%
%
%

\end{document}